\documentclass[epj]{svjour}
\usepackage{graphics}
\begin{document}
\hyphenation{com-bin-ed in-ter-ac-tio-on}
\headnote{    Preprint MPIH-V16-1999    }
\title{Estimates of the physical meson masses\\ 
       by an effective light-cone Hamiltonian}
%
%
\author{%
        Hans-Christian Pauli\inst
        }
\offprints{%
           Prof. H.C. Pauli, 
	   MPI f\"ur Kernphysik,
           Postfach 10 39 80, 
	   D-69029 Heidelberg,
           \\
           archive:  hep-th/9905xxx
        }
\institute{%
	   Max-Planck-Institut f\"ur Kernphysik, 
           Heidelberg, Germany.
           \email{pauli@mpi-hd.mpg.de}
           }
\date{29 Mai 1999} %
\authorrunning{H.C. Pauli}
\titlerunning{Meson masses}
\abstract{%
  A recent renormalization group analysis of
  the effective light-cone Hamiltonian yields 
  5 flavor masses and 2 mass shifts as free parameters.
  These are determined in the present note and
  used to calculate the masses of all 30 physical mesons
  (The topped mesons have been omitted).
  The agreement between experiment and theoretical estimate
  is quite satisfactory over the whole dynamical range,
  from the lightest pseudo-scalar mesons like the pion 
  up to the heaviest vector mesons like the upsilon.
  Evidence for an additive quark model is given and for 
  how the concept of isospin is realized in a gauge field theory
  such as QCD.
} 
\maketitle
%
%
\section{The mass formula}
\label{sec:1}

\begin{table}
\begin{tabular}{c|cccccc} 
     & $\overline u$ & $\overline d$ 
     & $\overline s$ & $\overline c$ & $\overline b$ \\ \hline
 $u$ &       &$\rho^+$&$K^{*+} $&$\overline D^{*0}$&$B^{*+}$\\ 
 $d$ &$\pi^-$&        &$K^{*0}$&$D^{*-}$&$ B^{*0}$\\ 
 $s$ &$K^-$  &$\overline K ^0$&      &$D_s^{*-}$&$B_s^{*0}$\\ 
 $c$ &$D^0$  &$D^+$&$D_s^+$&      &$B_c^{*+}$\\ 
 $b$ &$B^-$  &$\overline B ^0$&$\overline B_s ^0$&$B_c^{-}$&      \\ 
\end{tabular}
\caption{%
   The nomenclature of the `flavor-off-diagonal' physical mesons.
   The vector mesons are given in the upper, the pseudo-scalar mesons
   in the lower triangle.
}\label{tab:1}\end{table}
\begin{table}[b]
\begin{tabular}{c|rrrrrr} 
     & $\overline u$ & $\overline d$ 
     & $\overline s$ & $\overline c$ & $\overline b$ \\ \hline
 $u$ &      & 768  & 892  & 2007 & 5325 \\ 
 $d$ & 140  &      & 896  & 2010 & 5325 \\ 
 $s$ & 494  & 498  &      & 2110 &  --- \\ 
 $c$ & 1865 & 1869 & 1969 &      &  --- \\ 
 $b$ & 5278 & 5279 & 5375 &  --- &      \\ 
\end{tabular}
\caption{%
   The empirical masses of the flavor-off-diagonal physical mesons in MeV.
   The vector mesons are given in the upper, the pseudo-scalar mesons
   in the lower triangle.
}\label{tab:2}\end{table}
\begin{table}
\begin{tabular}{c|rrrrrr} 
     & $\overline u$ & $\overline d$ 
     & $\overline s$ & $\overline c$ & $\overline b$ \\ \hline
 $u$ &          & $^*$768 & 1002 & 2301 & 5696 \\ 
 $d$ & $^*$140  &         & 1002 & 2301 & 5696 \\ 
 $s$ & $^*$494  &     494 &      & 2535 & 5829 \\ 
 $c$ & $^*$1865 &    1865 & 2102 &      & 7227 \\ 
 $b$ & $^*$5278 &    5278 & 5512 & 6811 &      \\ 
\end{tabular}
\caption{%
   Same as Table~\protect{\ref{tab:2}} but for the calculated masses.
   The $^*$upperscript marks where shift and mass are determined.
}\label{tab:3}\end{table}
A recent renormalization group analysis of the effective 
light-cone Hamiltonian \cite{Pau99} culminates in the result that the
invariant mass-square eigenvalues $M^2$ of the physical
mesons are given by the quasi-static mass formula 
\begin{equation}
   M^2= (\overline m _{\bar q}+\overline m _{q'})^2 +
       2(\overline m _{\bar q}+\overline m _{q'})\ \overline s_\pm
.\label{eq:1}\end{equation}
In deriving this equation 
the dynamical effects of the binding energy proper 
have been omitted with an estimated error of about 10\%.
Some more background is given in Sections~\ref{sec:2} and~\ref{sec:5}.
The physical quark masses $\overline m _q$ and the mass shifts 
$\overline s_\pm$ are renormalization group invariants
and have to be fixed by experiment.
The mass shifts can have different values 
for the pseudo-scalar ($\overline s_-$) 
and the vector mesons ($\overline s_+$).
Their determination is one of the objects of this note.

The familiar nomenclature of the physical mesons is given 
in Table~\ref{tab:1}.
The empirical masses are taken from the data of the 
particle data group \cite{PDG98}.
These data do not yet include the topped mesons
and are compiled in Table~\ref{tab:2}. 
The following procedure was applied.
First, the up and the down mass were chosen equal,
$\overline m _{u}=\overline m _{d}=350$ MeV. 
Then the empirical masses of $\pi^+$ and $\rho^+$ were used 
to determine the mass shifts for the singlet and the triplet,
\begin{equation}
   \overline s_- = -336\ {\rm MeV}
   ,\quad{\rm and}\quad
   \overline s_+ = 71\ {\rm MeV}
,\end{equation}
respectively.
The remaining quark masses are obtained  
from the pseudo-scalar mesons with an up quark,
which exhausts all freedom in determining physical parameters. 
The resulting quark masses are given in Table~\ref{tab:4}. 
The remaining 13 off-diagonal pseudo-scalar and 
vector meson masses are calculated straightforwardly from 
Eq.(\ref{eq:1}), and compiled in Table~\ref{tab:3}.
Equivalently, but with slightly different results,
the quark masses could have been determined from
the off-diagonal vector mesons.

\section{The general problem}
\label{sec:2}

\begin{figure} 
  \resizebox{0.48\textwidth}{!}{%
  \includegraphics{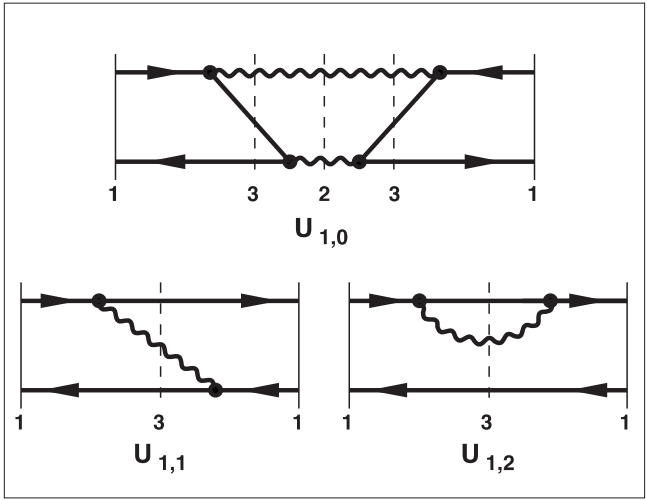} 
} \caption{
  The three graphs of the effective interaction in the 
  $q\bar q$-space.~---
  The lower two graphs correspond to the one-gluon-exchange
  interaction $U_{\rm OGE}$,
  the upper corresponds to the two-gluon-annihilation 
  interaction $U_{\rm TGA}$. 
  The figure is taken from Ref.\protect{\cite{Pau98}}
} \label{fig:1}
\end{figure}

Since the off-diagonal mesons are so well described, 
one wonders about the flavor-diagonal ones.

Let us review in short the general aspects. 
The full light-cone Hamiltonian for gauge theory
with its complicated many-body aspects is reduced in \cite{Pau98}
by the method of iterated resolvents 
to the effective Hamiltonian $H_{\rm eff}$ 
which by definition acts only 
in the Fock space of a single quark $q$ and 
a single anti-quark $\bar q$. It has essentially the two
contributions illustrated in Figure~\ref{fig:1}, {\it i.e.}
\begin{equation}
   H_{\rm eff} = T + U_{\rm OGE} + U_{\rm TGA}
.\label{eq:10}\end{equation}
The one-gluon-exchange interaction 
is combined with the kinetic energy $T$ into 
\begin{equation}
   H_{\rm OGE} = T  + U_{\rm OGE}
,\end{equation}
which can not change the flavor of the quark or the anti-quark.
It has been thoroughly investigated and 
renormalized in recent work \cite{Pau99,Pau98}.
As the figure shows, 
the effective two-gluon annihilation interaction $U_{\rm TGA}$
distroys a $q\bar q$-pair 
and creates another one with the same or an other flavor.
When addressing to the problem 
of diagonalizing the full effective interaction $H_{\rm eff}$, 
one must provide the Fock space for that. 
For three flavors, the complete Fock-space with any quark and any
anti-quark is given schematically in Table~\ref{tab:6}.
The matrix shown in this table visualizes the kernel of the effective
Hamiltonian as a matrix of block matrices. Each block matrix
represents a contribution from $H_{\rm eff}$. 
The symbol $E$ stands for $H_{\rm OGE}$. 
The blocks labelled with $A$ indicate $U_{\rm TGA}$. 
Different indices refer different numbers inside the block.
The block $E_2=\langle u\bar s \vert H_{\rm OGE}\vert u\bar s\rangle$,
for example, is the same as 
$E'_2=\langle s\bar u \vert H_{\rm OGE}\vert s\bar u\rangle$
since charge exchange does not change the eigenvalues.
Most blocks correspond to a zero matrix.
The way these zero matrices arrange themselves in the table
is peculiar to QCD. This demonstrates 
that most of the Hamiltonian is reducible 
and that one can diagonalize blockwise. 
The blockwise diagonalization of the off-diagonal mesons 
in Tables~\ref{tab:2} and \ref{tab:3}
survives thus even if the annihilation interaction is included.

The last three rows and columns in Table~\ref{tab:6} however show, 
that the annihilation mixes the flavors in the charge-neutral
flavor-diagonal mesons. 
The QCD-flavor mixing for 3 flavors is thus the $3\times3$
block matrix in Table~\ref{tab:6}.
For $N$ flavors one has correspondingly a $N\times N$ matrix.

\begin{table}
\begin{tabular}{c|ccc ccc ccc} 
     & $u\overline d$ & $u\overline s$ & $d\overline s$ 
     & $d\overline u$ & $s\overline u$ & $s\overline d$ 
     & $u\overline u$ & $d\overline d$ & $s\overline s$ \\ \hline
 $u\overline d$ &$E_1$ & 0    & 0    & 0    & 0    & 0    & 0    & 0    & 0   \\ 
 $u\overline s$ & 0    &$E_2$ & 0    & 0    & 0    & 0    & 0    & 0    & 0   \\
 $d\overline s$ & 0    & 0    &$E_3$ & 0    & 0    & 0    & 0    & 0    & 0   \\ 
 $d\overline u$ & 0    & 0    & 0    &$E'_1$& 0    & 0    & 0    & 0    & 0   \\ 
 $s\overline u$ & 0    & 0    & 0    & 0    &$E'_2$& 0    & 0    & 0    & 0   \\ 
 $s\overline d$ & 0    & 0    & 0    & 0    & 0    &$E'_3$& 0    & 0    & 0   \\ 
 $u\overline u$ & 0    & 0    & 0    & 0    & 0    & 0    &$e_4$ &$A_5$ &$A_6$\\ 
 $d\overline d$ & 0    & 0    & 0    & 0    & 0    & 0    &$A_5$ &$e_7$ &$A_8$\\ 
 $s\overline s$ & 0    & 0    & 0    & 0    & 0    & 0    &$A_6$ &$A_8$ &$e_9$\\ 
\end{tabular}
\caption{%
   The kernel of the effective Hamiltonian is displayed as
   a block matrix to illustrate the falvor mixing in QCD.
   The abbreviation $e_i=E_i+A_i$ is used by obvious reasons.
}\label{tab:6}\end{table}
How can one include the annihilation graph $U_{\rm TGA}$, 
at least approximately?
The numerators in the annihilation graph of Figure~\ref{fig:1}
must yield a Lorentz scalar like the current coupling
$[\overline u (q)\gamma^\mu v (q)]\, 
 [\overline v(q')\gamma_\mu u(q')]$.
Because of our Dirac-spinor convention \cite{BroPauPin98,LepBro80} 
one can thus anticipate that 
$U_{\rm TGA} \sim \overline m_{q} \overline m_{q'} $.
The propagator in the two-gluon intermediate state
was proposed in \cite[Eq.(69)]{Pau98} to obey
${\overline M_g^2 - \frac{1}{2} (M_{qq}^2+M_{q'q'}^2)} ]^{-1}$ 
with the glueball mass $\overline M_g$.
As reasonable estimate one can use $\overline M_g\simeq 1.4\ {\rm GeV}$. 
The matrix elements of $U_{\rm TGA}$ are therefore parametrized as
\begin{equation}
    \langle{q\bar q}\vert U_{\rm TGA}\vert{q'\bar q'}\rangle
    = b\  %
    \frac{\overline m_{q} \overline m_{q'}}{\overline M_g^2 - \frac{1}{2}
    \left(M_{q\bar q}^2+M_{q'\bar q'}^2\right)} 
.\label{eq:5}\end{equation}
In the present context, the coefficient $b$ is subject to be 
determined by a fit to the experiment. 
It can have either sign, 
and can be different for pseudo-scalar and vector mesons. 

The nature and physical interpretation of the parameter
$b$ is very different from for example $\overline s_\pm$.
The latter is a physical (renormalization) constant,
while Eq.(\ref{eq:5}) and the coefficients $b_S$ and $b_V$
are kind of empirical constraints on a future and more accurate 
analysis of the matrix elements of $U_{TGA}$.

\section{Flavor SU(2)}
\label{sec:3}

\begin{table}
\begin{tabular}{||cr||rr||rr||} 
\multicolumn{2}{||c||} {quarks}
               &\multicolumn{2}{c||} {pseudo-scalar bosons}
	                      &\multicolumn{2}{c||} {vector bosons} \\
   \rule[-1em]{0mm}{1em}
   $q$ & $\overline m_q$ 
	       & Eq.(\protect{\ref{eq:1}}) & SU(2) 
	                     & Eq.(\protect{\ref{eq:1}}) & SU(2) \\
  \hline
   \rule[1em]{0mm}{0.5em}
  $u$  &   350 &  140 &  140 &  768 &  768 \\ 
  $d$  &   350 &  140 &  549 &  768 &  782 \\ 
  $s$  &   583 &  760 &      & 1236 &      \\ 
  $c$  &  1881 & 3410 &      & 3833 &      \\ 
  $b$  &  5275 &10208 &      &10620 &      \\ 
\end{tabular}
\caption{%
   Column 1 and 2 compile the symbols and the masses 
   of the quarks as obtained from the fit to the 
   flavor-off-diagonal pseudo-scalar mesons.--- 
   Column 3 gives the contribution from the one-gluon exchange
   interaction to the pseudo-scalar mesons, Eq.(\protect{\ref{eq:1}}). 
   Column 4 compiles the results for SU(2) flavor mixing.
   Column 5 and 6 give the corresponding numbers 
   for the vector mesons.
}\label{tab:4}\end{table}
Consider first 2 flavors with equal masses 
$\overline m_u=\overline m_d$ for the charge neutral pseudo-scalar mesons.
The flavor-mixing matrix is
\begin{equation}
 H_{\rm eff} = 
 \bordermatrix{%
         & u\bar u           & d\bar d           \cr 
 u\bar u & a + M^2_{u\bar u} & a \cr 
 d\bar d & a & a + M^2_{u\bar u} \cr 
}.\label{2eq:20}\end{equation}
Because of the equal quark masses, the Hamiltonian $H_{\rm OGE}$
is identical in the $u\bar u$ and the $d\bar d$ sector with
identical eigenstates $\vert\Psi_{q\bar q;i}\rangle$
and eigenvalues $M_{q\bar q;i}^2$. The matrix elements
of $U_{\rm TGA}$ are therefore all equal, 
$a =\langle{u\bar u}\vert U_{\rm TGA}\vert{u\bar u}\rangle =
    \langle{u\bar u}\vert U_{\rm TGA}\vert{d\bar d}\rangle =
    \langle{d\bar d}\vert U_{\rm TGA}\vert{d\bar d}\rangle  
$. 
One can thus diagonalize state after state, and can interpret 
Eq.(\ref{2eq:20}) as a simple 2 by 2 matrix 
(instead of a 2 by 2 {\em block matrix}), with
$M^2_{u\bar u}$ given by Eq.(\ref{eq:1}). 

The diagonalization of 
$H_{\rm eff} \vert\Phi_{i}\rangle 
 = M_{i}^2 \vert\Phi_{i}\rangle$ 
is easy. The two eigenstates are
\begin{equation}
   \vert\Phi_{1}\rangle = \frac {1}{\sqrt{2}} 
   \pmatrix{%
              \vert u\bar u \rangle \cr
              \vert d\bar d \rangle \cr
   }%
   ,\ \vert\Phi_{2}\rangle = \frac {1}{\sqrt{2}} 
   \pmatrix{%
              \phantom{-}\vert u\bar u \rangle \cr
              -          \vert d\bar d \rangle \cr
   }%
,\end{equation}
and are associated with the eigenvalues
\begin{equation}
   M_{1}^2 = M^2_{u\bar u} + 2a  
   ,\qquad 
   M_{2}^2 = M^2_{u\bar u}  
.\end{equation}
The coherent state $\Phi_{1}$ picks up all the strength.
Since the eigenvalue of the incoherent state $\Phi_{2}$ is 
identical with the unperturbed value $M^2_{u\bar u}$, 
it is reasonable to interpret it as the $\pi^0$, 
which combines with the $\pi^+$ and the $\pi^-$
into a mass degenerate triplet of {\em isospin} $I=1$. 
The coherent state can be fitted to the $\eta$ 
which determines $b\equiv b_S$.
The outcome of this fit is reported in Table~\ref{tab:4}.
The same procedure can be applied to the vector mesons,
correspondingly, with the resulting fit values   
\begin{equation}
   b _S = (1494{\rm\ MeV})^2,\quad 
   b _V = ( 348{\rm\ MeV})^2
.\end{equation}
The results for the vector mesons are compiled in Table~\ref{tab:4}.

\section{Flavor SU(3)}
\label{sec:4}
\begin{table}
\begin{tabular}{||crr||crr||} 
\multicolumn{3}{||c||} {pseudo-scalar bosons}
	                   &\multicolumn{3}{c||} {vector bosons} \\
  \rule[-1em]{0mm}{1em}
  name     &  exp &  SU(3) & name       &  exp & SU(3) \\ \hline
  \rule[1em]{0mm}{0.5em}
  $\pi^0$  &  135 &    140 & $\rho^0$   &  768 &    768 \\ 
  $\eta $  &  549 &$^*$549 & $\omega$   &  782 &$^*$782 \\ 
  $\eta'$  &  958 &    981 & $\Phi  $   & 1019 &   1425 \\ 
  $\eta_c$ & 2980 &   3394 & $J/\Psi $  & 3097 &   3848 \\ 
  $\eta_b$ &  --- &  10204 & $\Upsilon$ & 9460 &  10625 \\ 
\end{tabular}
\caption{%
   Column 1 and 2 compile the names and the masses 
   of the pseudo-scalar mesons.  
   Column 3 gives the theoretical estimate for SU(3) 
   flavor mixing.
   Column 4, 5 and 6 give the corresponding numbers 
   for the vector mesons.---
   The $^*$star indicates which masses have been fitted.
}\label{tab:5}\end{table}
Next, consider 3 flavors with equal masses 
$\overline m_s=\overline m_d=\overline m_u$. 
The flavor mixing matrix is then
\begin{equation}
 H_{\rm A} = 
 \bordermatrix{%
         & u\bar u & d\bar d & s\bar s \cr 
 u\bar u & a + M^2_{u\bar u} & a & a \cr 
 d\bar d & a & a + M^2_{u\bar u} & a \cr 
 s\bar s & a & a & a + M^2_{u\bar u} \cr 
},\label{3eq:20}\end{equation}
with $H_{\rm eff}\equiv H_{\rm A} $.
Up to an obvious normalization constant
the three eigenstates are 
\begin{equation}
   \vert\Phi_{1}\rangle =  
   \pmatrix{%
              \vert u\bar u \rangle \cr
              \vert d\bar d \rangle \cr
              \vert s\bar s \rangle \cr
   }%
  ,\    \vert\Phi_{2}\rangle =  
   \pmatrix{%
             \phantom{-}\vert u\bar u \rangle \cr
                      - \vert d\bar d \rangle \cr
                      0 \vert s\bar s \rangle \cr
   }%
   ,\ \vert\Phi_{3}\rangle = 
   \pmatrix{%
             \phantom{- 2}\vert u\bar u \rangle \cr
             \phantom{- 2}\vert d\bar d \rangle \cr
                      - 2 \vert s\bar s \rangle \cr
   }%
\end{equation}
and have the eigenvalues 
\begin{equation}
   M_{1}^2 = M^2_{u\bar u} + 3a  
   ,\quad 
   M_{2}^2 = M^2_{u\bar u}  
   ,\quad 
   M_{3}^2 = M^2_{u\bar u}  
.\end{equation}
The coherent state picks up all the strength again. 
The eigenvalues of the remaining two states are again 
the unperturbated ones, which now are degenerate. 
State $\Phi_{2}$ can again be interpreted as 
the eigenstate for the charge neutral $\pi^0$ and the mass of
the coherent state $\Phi_{1}$ can be fitted with the $\eta$.
But then state $\Phi_{3}$ is degenerate in mass with the $\pi^0$.

Obviously, one cannot abstract from the appreciable mass difference
of up and strange quark.
Including the different masses yields $H_{\rm eff}$ which
for convenience is split up like 
$H_{\rm eff}= H_{\rm A} + H_{\rm C} $,
with $H_{\rm A}$ given in Eq.(\ref{3eq:20})
and thus $H_{\rm C}$ by
\begin{equation}
 H_{\rm C} = 
 \bordermatrix{%
         & u\bar u & d\bar d & s\bar s \cr 
 u\bar u & 0   & 0   & C_1 \cr 
 d\bar d & 0   & 0   & C_2 \cr 
 s\bar s & C_1 & C_2 & C_d \cr 
}.\end{equation}
Because of $\overline m_u=\overline m_d$ one has $C_1=C_2$ and 
\begin{eqnarray}
   C_1 &=& \langle {u\bar u}\vert U_{\rm TGA}\vert{s\bar s}\rangle -
           \langle {u\bar u}\vert U_{\rm TGA}\vert{u\bar u}\rangle 
,\\    
   C_d &=& \langle {s\bar s}\vert U_{\rm TGA}\vert{s\bar s}\rangle -
           \langle {u\bar u}\vert U_{\rm TGA}\vert{u\bar u}\rangle +
	   M^2_{s\bar s} - M^2_{u\bar u} 
.\end{eqnarray}
The $3\times 3$ matrix can not be diagonalized analytically,
at least not in a simple way. 
Therefore, $H_{\rm C}$  is included 
by calculating the expectation values of $H_{\rm C}$
with the eigenstates of $H_{\rm A}$, {\it i.e.} 
\begin{eqnarray}
   \langle\Phi_{1}\vert H_{\rm C}\vert\Phi_{1}\rangle &=& 
   \frac {1}{3} (C_d + 2C_1+ 2C_2)
,\nonumber\\    
   \langle\Phi_{2}\vert H_{\rm C}\vert\Phi_{2}\rangle &=& 0
,\\    
   \langle\Phi_{3}\vert H_{\rm C}\vert\Phi_{3}\rangle &=&
   \frac {2}{3} (C_d - C_1- C_2)
.\nonumber\end{eqnarray}
Adding these to the eigenvalues of $H_{\rm A}$ gives
\begin{eqnarray}
   M_{1}^2 &=& M^2_{u\bar u} 
   + \langle\Phi_{1}\vert H_{\rm C}\vert\Phi_{1}\rangle + 3a
,\nonumber\\    
   M_{2}^2 &=& M^2_{d\bar d} 
   + \langle\Phi_{2}\vert H_{\rm C}\vert\Phi_{2}\rangle 
,\\    
   M_{3}^2 &=& M^2_{s\bar s} 
   + \langle\Phi_{3}\vert H_{\rm C}\vert\Phi_{3}\rangle
.\nonumber\end{eqnarray}
The coefficient $b=b_S$ is fixed for the pseudo-scalar mesons 
by requiring $M_{1}=M_{\eta}$.
For the vector mesons, 
the coefficient $b=b_V$ is fixed by requiring $M_{1}=M_{\omega}$.
The result is
\begin{equation}
   b_S= (539{\rm\ MeV})^2
   ,\quad{\rm and}\quad
   b_V= (652{\rm\ MeV})^2
.\end{equation}
Finally, the physical mass of the neutral charmed pseudo-scalar meson 
is corrected according to Eq.(\ref{eq:5}), like
\begin{equation}
   \left(M _{c\bar c}^2\right)_{physical} = 
   M _{c\bar c}^2 + b_S \ \frac{\overline m_s ^2}
   {\overline M _{gg}^2 - M _{c\bar c}^2}
,\label{2eq:11}\end{equation}
in order to account for the annihilation channel approximatively.
The flavor-diagonal $b$-mesons and 
vector mesons are treated correspondingly.
The results are compiled in Table~\ref{tab:5}.

\section{Discussion and conclusions}
\label{sec:5}
In this necessarily short note all known hadron masses
from the lightest ones like the pion
up to the heaviest ones like the upsilon
are described within the same model using the same parameters.
Its ingredients are based on the effective Hamiltonian, 
down to which the genuine many-body aspects of gauge field theory 
and the full light-cone Hamiltonian for QCD
are reduced by the method of iterated resolvents,
and its subsequent analysis with the renormalization group.

The so obtained Eq.(\ref{eq:1}) is the quasi-static limit of
the eigenvalues of the full effective Hamiltonian.
For sufficiently large quark masses one can expand 
Eq.(\ref{eq:1}) as a power series in
$\overline s_\pm/(\overline m _{\bar q}+\overline m _{q'}) $,
\begin{equation}
   M\simeq \overline m _{\bar q}+\overline m _{q} + \overline s_\pm
   + \cdots
,\label{eq:2}\end{equation}
which to leading order yields an {\em  additive quark model}.

Isospin is not a dynamic symmetry of QCD. 
In the present light-cone approach to gauge theory with an
effective interaction
isospin arises because the mass of the strange quark is similar to
but different from the mass of up and down.
Flavor-SU(3) is an approximate symmetry.
The light-cone approach can even explain why the phenomenological 
flavor-SU(3) symmetry works better than flavor-SU(4) or SU(5):
the large mass of the heavy mesons dominates the flavor-mixing
matrix so strongly that the symmetry induced by the annihilation
interaction is destroyed. The model predicts however that
all flavors are mixed in the charge neutral mesons.
The $\pi^+$ and the $\pi^-$ are obtained by fixing the quark mass and  
the mass shift. The third member of the isospin triplet, 
the $\pi^0$ is then obtained for free.
The $\eta$ is obtained by fitting the parameter $b_S$ which accounts
for the two-gluon annihilation interaction in an approximate way.
The $\eta'$ is then obtained for free with an accuracy of about 5\%.
The present work contributes thus to the $\eta$-$\eta'$ 
puzzle \cite{BurGoi97}
and exposes an accuracy comparable to state-of-art
lattice gauge calculations \cite{Kil97}.

The empirical and the calculated masses for the flavor-off-diagonal 
mesons show strong correlation.  
The largest discrepancy of roughly 15\% is observed for the charmed
vector mesons. On the average they agree with each other 
within 5\% or less. 
It can be expected that the agreement gets even better
in more accurate work where the dynamics is included.

To our recollection there is no other model 
which covers the whole range of flavored hadrons
with the same set of parameters.
This includes the phenomenological models \cite{DonGolHol92}, 
where the light flavors are very successfully described by
the Nambu-Jona-Lasinio model and its derivatives. 

\begin{acknowledgement}
It is a pleasure to thank 
Susanne Bielefeld for checking the calculations.
\end{acknowledgement}
\bibliography{20bib}
%

\end{document}